\documentclass{vgtc}                          

\graphicspath{{figures/}{pictures/}{images/}{./}} 

\usepackage{times}                     

\usepackage{tabu}                      
\usepackage{booktabs}                  
\usepackage{lipsum}                    
\usepackage{mwe}                       
\usepackage{amsmath}
\usepackage{mathptmx}                  
\usepackage{kotex}
\usepackage{listings}
\usepackage{fancyvrb}

\newsavebox{\FVerbatimbox}
\newlength{\FVerbatimwidth}
\AtBeginDocument{%
  \sbox0{\ttfamily X}%
  \setlength{\FVerbatimwidth}{60\wd0}%
  \addtolength{\FVerbatimwidth}{2\fboxsep}
}

\onlineid{1023}

\vgtccategory{Research}

\vgtcinsertpkg

\title{Can VLMs Assess Similarity Between Graph Visualizations?}

\author{
Seokweon Jung \thanks{e-mail: \{swjung, hj, jmrhee\}@hcil.snu.ac.kr}  \quad  
Hyeon Jeon\footnotemark[1]  \quad
Jeongmin Rhee\footnotemark[1] \quad 
Jinwook Seo\thanks{e-mail: jseo@snu.ac.kr}  \\ 
Seoul National University \\
}

\teaser{
  \centering
  \includegraphics[width=\linewidth]{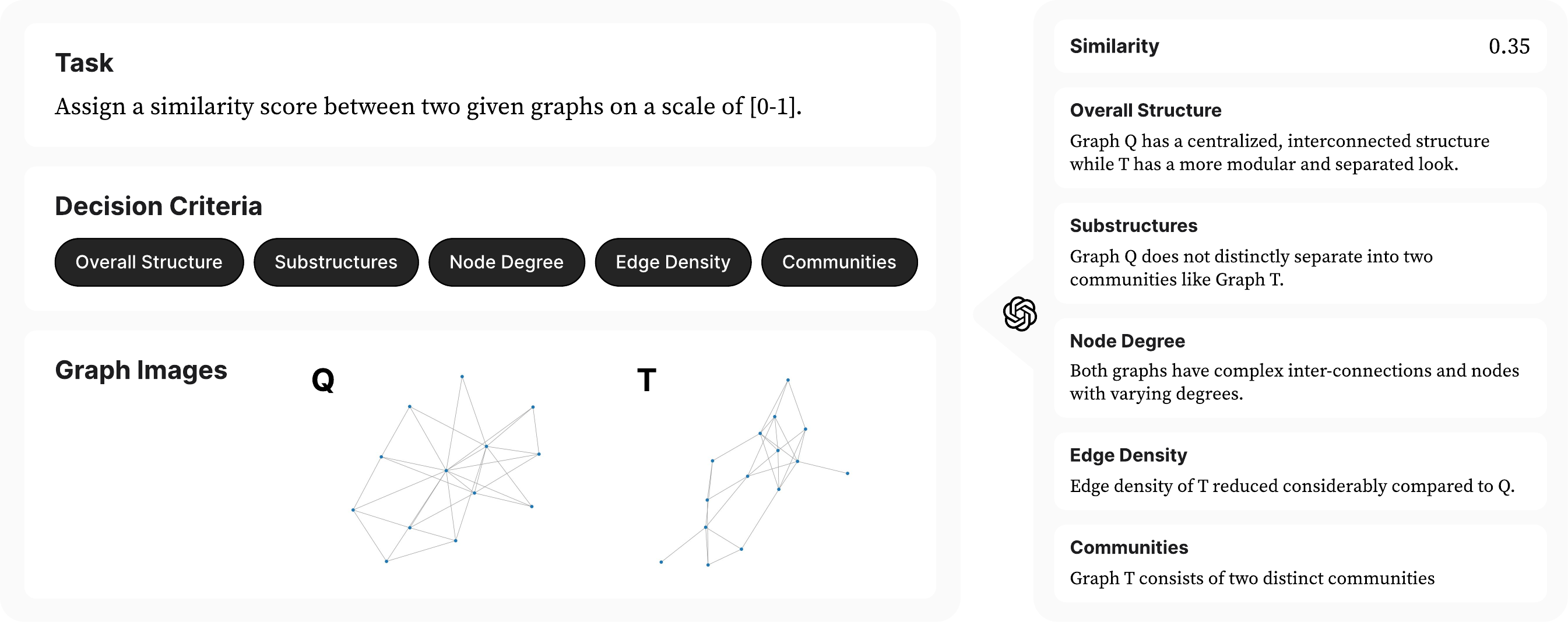}
  \caption{An example of graph similarity assessing process with VLM (gpt-4o). It receives two node-link diagram images, and it is prompted to assess a similarity score and provide reasonings according to decision criteria. The result supports the fact that gpt-4o can capture the differences between images of two graph visualizations.}
  \label{fig:teaser}
}

\abstract{Graph visualizations have been studied for tasks such as clustering and temporal analysis, but how these visual similarities relate to established graph similarity measures remains unclear. In this paper, we explore the potential of Vision Language Models (VLMs) to approximate human-like perception of graph similarity. We generate graph datasets of various sizes and densities and compare VLM-derived visual similarity scores with feature-based measures. Our findings indicate VLMs can assess graph similarity in a manner similar to feature-based measures, even though differences among the measures exist. In future work, we plan to extend our research by conducting experiments on human visual graph perception.
} 


\begin{document}


\firstsection{Introduction}
\maketitle
Graph visualizations for exploring the visual similarity between graphs have been investigated from various perspectives, such as clustering similar graphs~\cite{adamotif} or analyzing temporal evolution~\cite{monetexplorer}. 
In these studies, mathematically or computer science-derived graph similarity measures or graph distances serve as important indicators for evaluating similarities among graphs. 

However, although measuring the similarities between individual graphs has been studied for over fifty years~\cite{emmert2016survey}, the relationship between visual perception of graph similarity and these measures remains unexplored. In other visualization domains, such as scatterplots, researchers have attempted to use experimentally measured visual perceived similarity as a baseline for similarity evaluation.
If we investigate the correlation between visual similarity and various graph similarity measures, they could provide a valuable baseline for graph comparison tasks.

Prior to undertaking extensive human studies on graph similarity perception, we aim to simulate this perceptual ability using Vision Language Models (VLMs). Although it is known that VLMs’ performance in interpreting visualizations is currently limited~\cite{vlm-blind}, rapid advancements have spurred various attempts to leverage VLMs for visual interpretation~\cite{channel-effectivness, shpark-search}. 

In this study, we employ gpt-4o to assess the visual similarity between graph visualizations. We generate graph datasets with diverse sizes and densities~\cite{survey-emprical-graphvis}. We then compare these results with feature-based graph similarity measures to investigate the following: 1) under which size and density conditions VLMs can effectively interpret graph similarity, and 2) how well the VLMs’ interpretation correlates with traditional feature-based measures.

\section{Background and Related Work}

Traditional mathematical and computer science-based graph similarity measures can be broadly categorized into two groups: Known Node Correspondence (KNC) algorithms, which are applicable to deterministic networks, and Unknown Node Correspondence (UNC) algorithms, which are suited for random networks~\cite{emmert2016survey}.
Similarity measures of KNC, such as graph edit distance and Jaccard similarity, assume that the node labels are known. In contrast, the UNC similarities compute graph similarity by extracting features beyond node labels, ranging from low-level attributes like degree distributions to high-level substructures~\cite{tantardini2019survey}.

Besides these similarity measures, the visualization community has long pursued tasks that assess similarity through comparisons between various static graphs~\cite{adamotif} or by analyzing evolving patterns in dynamic graphs over time~\cite{monetexplorer}. However, although these studies have conducted visual analyses of graphs, the basic comparisons have relied on the aforementioned similarity measures, and there is a lack of quantitative evaluation regarding how well visual differences are perceived. While attempts have been made to perceive changes in properties by varying the size, density, and layout of graph visualizations~\cite{perception-of-graphproperties}, a direct evaluation of the similarity between two graphs has not yet been performed.

In this study, we focus on the most general and simple graph visualizations: node-link diagrams of random, undirected, and unweighted networks with force-directed layouts with UNC similarity measures. By varying the size and density, we assess the similarity perceived in graph visualizations using VLMs.

\section{Experiment Design}

\subsection{Dataset}
We conducted experiments using the node-link diagram with force-directed layout, the most commonly used representation in graph visualization~\cite{perception-of-graphproperties}. To simplify the problem situation, we limited the graph to be unweighted and undirected.

Based on a survey paper that empirically examined the size and density of graph data used in graph visualization experiments, we categorized the graphs by four sizes and three linear density classes~\cite{survey-emprical-graphvis}.
In total, there were 12 types of size-density combinations. For each combination, eight graphs were generated per synthetic graph generation algorithm: GNM (random), BBA (centralized), NWS (circular), and SBM (multi-communities), resulting in a total of 384 images and 190,464 pairwise similarity sets.
All generated data are a single connected component, as we aimed to facilitate one-to-one comparisons between graphs.

\begin{table}[]
    \centering
        \caption{Graph similarity measures based on graph attributes. \(|V|\): number of vertices, \(|E|\): number of edges, \(JSD\): Jensen-Shannon divergence, \(Lv\): Louvain community detection.}
    \begin{tabular}{|m{2.7cm} | m{0.5cm} | m{3.9cm}|}
        \toprule
         \footnotesize \textbf{Measure} & \footnotesize \textbf{Abb} & \footnotesize \textbf{Formula} \\
         \midrule
         \midrule
         \footnotesize Size  & \textit{S} &  \(1 - ||V_1|-|V_2||/max(|V_1|, |V_2|)\)\\
         \midrule
         \footnotesize Density (Linear) & \textit{D}& \(1-\frac{||E_1|/|V_1|- |E_2|/|V_2||}{max(|E_1|/|V_1|, |E_2|/|V_2|)}\) \\
         \midrule
         \footnotesize Node degree& \textit{Nd} & \(1 - JSD(Nd(G1), Nd(G2))\) \\ 
         \midrule
         \footnotesize Clustering coefficient  & \textit{Cc} & \(1 - JSD(Cc(G1), Cc(G2))\) \\
          \midrule
        \footnotesize Betweenness centrality  & \textit{Bc} & \(1 - JSD(Bc(G1), Bc(G2))\)\\
          \midrule
         \footnotesize Community distribution  & \textit{Cm} & \(1 - JSD(Lv(G1), Lv(G2))\)\\
         \bottomrule
    \end{tabular}
    \label{tab:algorithms}
\end{table}

\begin{figure}[tb]
 \centering 
 \includegraphics[width=\columnwidth]{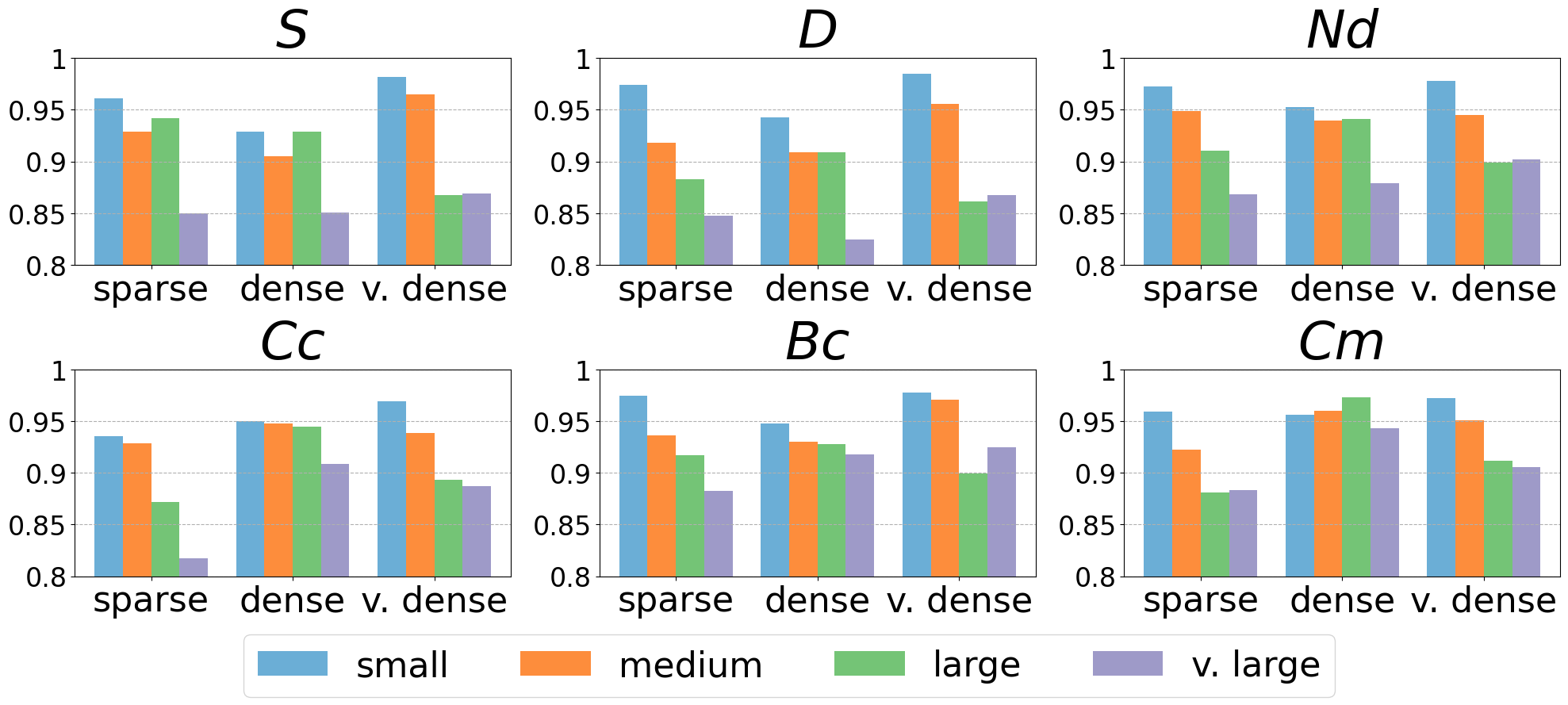}
\vspace{-0.5cm}
 \caption{Pearson correlation coefficient between six similarity measures in ~\autoref{tab:algorithms} and gpt-4o.}
 \vspace{-0.5cm}

 \label{fig:result}

\end{figure}

\subsection{VLM Model Prompting}

We utilized the \texttt{gpt-4o-2024-08-06} model\footnote{\url{https://platform.openai.com/docs/models/gpt-4o}} to assess the performance of VLM's graph similarity perception.
We chose the model because it is among the currently available VLMs that offer fast response times and high performance. As shown in ~\autoref{fig:teaser}, We provided the VLM with two images and prompted it to output the similarity between the two graphs as a value between 0 and 1. Additionally, we requested that the model explain the basis for the similarity rating using five criteria: (1) overall structure, (2) specific substructures or repeating patterns, (3) node degrees, (4) edge density, and (5) community distribution. Furthermore, if there were any other features that could serve as a basis for the similarity assessment, we asked for those to be elaborated upon as well. 

\subsection{Graph Similarity Measures}
Among the various graph similarity measures applicable to random networks, we adopted basic feature-based measures. Not only size (1, 3) and density (4), node degree (3), clustering coefficient (2, 4), betweenness centrality (1, 2), and community size (5) are selected to reflect decision criteria~\autoref{tab:algorithms}.

\section{Results}
Using the Pearson correlation coefficient, we analyzed the relationship between the similarity determined by the VLM and the six feature-based graph similarities. The key findings are as follows:
\vspace{-0.2cm}
\begin{enumerate}
    \item Across all size and density combinations, correlations were consistently strong (≥ 0.8).
    \vspace{-0.2cm}
    \item Under the same density condition, correlation decreases as graph size increases.
    \vspace{-0.2cm}
    \item Under the same size condition, there was no clear trend in correlation even as density increased.
    \vspace{-0.2cm}
    \item High-level features, such as \textit{Cm} and \textit{Bc}, exhibit a higher correlation with the VLM in larger, denser graphs.
\end{enumerate}

\section{Conclusion}
We explore the potential of VLMs ability to assess graph similarity by comparing it with feature-based similarity measures. 
Our experiments indicate that while the VLM’s judgments are generally similar to feature-based graph similarity measures. However, the observed differences among the measures suggest that a more refined measure could potentially align more closely with the VLM’s assessments. In future work, we aim to examine whether similar differences emerge in human visual perception and to identify which measure best approximates human perception.
\bibliographystyle{abbrv-doi}

\bibliography{template}

\begin{thebibliography}{1}

\bibitem{emmert2016survey}
F.~Emmert-Streib, M.~Dehmer, and Y.~Shi.
\newblock Fifty years of graph matching, network alignment and network comparison.
\newblock {\em Inf. Sci. (Ny)}, 346-347:180--197, June 2016.

\bibitem{monetexplorer}
S.~Jung, D.~Shin, H.~Jeon, K.~Choe, and J.~Seo.
\newblock Monetexplorer: A visual analytics system for analyzing dynamic networks with temporal network motifs.
\newblock {\em IEEE Transactions on Visualization and Computer Graphics}, 2023.

\bibitem{channel-effectivness}
S.~Lee, M.~Chang, S.~Park, and J.~Seo.
\newblock Assessing graphical perception of image embedding models using channel effectiveness.
\newblock In {\em 2024 IEEE Visualization and Visual Analytics (VIS)}, pp. 226--230. IEEE, 2024.

\bibitem{shpark-search}
S.~Park, Y.~Song, S.~Lee, J.~Kim, and J.~Seo.
\newblock Leveraging multimodal llm for inspirational user interface search.
\newblock {\em arXiv preprint arXiv:2501.17799}, 2025.

\bibitem{vlm-blind}
P.~Rahmanzadehgervi, L.~Bolton, M.~R. Taesiri, and A.~T. Nguyen.
\newblock Vision language models are blind.
\newblock In {\em Proceedings of the Asian Conference on Computer Vision}, pp. 18--34, 2024.

\bibitem{perception-of-graphproperties}
U.~Soni, Y.~Lu, B.~Hansen, H.~C. Purchase, S.~Kobourov, and R.~Maciejewski.
\newblock The perception of graph properties in graph layouts.
\newblock {\em Computer Graphics Forum}, 37(3):169--181, 1~June 2018.

\bibitem{tantardini2019survey}
M.~Tantardini, F.~Ieva, L.~Tajoli, and C.~Piccardi.
\newblock Comparing methods for comparing networks.
\newblock {\em Sci. Rep.}, 9(1):17557, 26~Nov. 2019.

\bibitem{survey-emprical-graphvis}
V.~Yoghourdjian, D.~Archambault, S.~Diehl, T.~Dwyer, K.~Klein, H.~C. Purchase, and H.-Y. Wu.
\newblock Exploring the limits of complexity: A survey of empirical studies on graph visualisation.
\newblock {\em Vis. Inform.}, 2(4):264--282, Dec. 2018.

\bibitem{adamotif}
H.~Zhou, P.~Lai, Z.~Sun, X.~Chen, Y.~Chen, H.~Wu, and Y.~Wang.
\newblock {AdaMotif}: Graph simplification via adaptive motif design.
\newblock {\em IEEE Trans. Vis. Comput. Graph.}, PP(99):1--11, 10~Sept. 2024.

\end{thebibliography}
\end{document}